\documentclass[10pt]{article}
\usepackage{latexsym}
\usepackage{eufrak}
\usepackage{euscript}
\usepackage{graphicx}
\usepackage[dvips]{color}

\title{A brief overview and some comments on the weak measurement protocol}

\author{P. N. Kaloyerou\\ \mbox{}\\The University of Zambia,  School of Natural Sciences, \\  Department of Physics, Lusaka 10101,  Zambia\\ and\\ Wolfson College, Linton Road, Oxford OX2 6UD, UK 
              \footnote{email address: pan.kaloyerou@wolfson.ox.ac.uk}}

\begin{document}

\maketitle


\begin{abstract}
\mbox{}\\
The main purpose of this short article is to give a brief overview of the development of the very interesting weak measurement protocol. I add some comments relating to the reality  of weak values, and also comment on the allowed values of observables between measurements.
\mbox{}\\ \mbox{}\\
keywords: {quantum mechanics, weak measurements, Bohm and de Broglie and causal interpretation}
\mbox{}\\  \mbox{}\\
PACS NO: 03.65.Ta
\end{abstract}

\section{Brief overview  of the development of the weak measurement protocol\label{bro}}
The weak measurement protocol \footnote{The weak measurement protocol consists of a weak interaction followed by a strong interaction followed by a final irreversible (in the statistical sense), amplified detection. Commonly used  terminology refers to the weak and strong interactions as measurements. Though, perhaps, it is well understand, it is still worth noting that  neither the weak nor the strong interaction fulfill even a nonrigorous statistical definition of a measurement and should, strictly, be referred to as interactions. The final irreversible detection certainly satisfies the conditions for a measurement, and hence, so does the whole weak measurement protocol. A discussion  of what constitutes a measurement may be found in reference  \cite{K16}. However, since the terminology is widespread we will also use ``measurement" to refer to the weak and strong interactions.},  (or just  `weak measurement' for short) has gained fairly wide prominence since its early, rather abstract, beginnings and has found important practical applications. Here, I give a brief overview of the weak measurement protocol introduced by Aharonov et al  and offer some comments.

Early ideas that eventually led to the weak measurement protocol began in a 1964 article by Aharonov, Bergmann and Lebowitz \cite{ABL64}. In this article the idea of time asymmetry in quantum mechanics was examined. The authors suggested that time symmetric ensembles could be obtained by first measuring an observable represented by the operator\footnote{In what follows we will use the term observable to also refer to the operator representing the observable. The same letter representing the observable will, therefore, also be used  to represent the operator.} $A$   at time $t_0$ (preselection) and then, at a later time $t_f$, measuring  an observable $B$ (postselection), where $A$ and $B$ may or may not commute.  Preselection puts all members of the ensemble into an eigenstate $| A,a \rangle$, with eigenvalue $a$, of the operator $A$, while postselection places the ensemble in an eigenstate  $|B,b\rangle$, with eigenvalue $b$, of operator $B$.  They also derived a formula for the probability  of finding eigenvalues  $d_1, d_2,\ldots d_k$ of a set of  observables  $D_1, D_2,\ldots D_k$, where the observables are either measured in sequence or simultaneously between the preselection measurement and the postselection measurement. They operators $D_1, D_2,\ldots  D_k$ may or may not commute. For the case of a complete   set of commuting observables, collectively represented by  $C$, 
 with eigenvalues collectively represented by $c$, the formula  for the probability $P(c_j)$ of obtaining eigenvalues $c_j$ becomes
\begin{equation}
P(c_j)=\frac{|\langle A,a|C,c_j \rangle|^2 |\langle C,c_j |B,b \rangle|^2}{\sum_i|\langle A,a|C,c_i \rangle|^2 |\langle C,c_i |B,b \rangle|^2}.\label{ABLP}
\end{equation}

Concerning the time-asymmetry of quantum mechanics they concluded that quantum mechanics is not time symmetric despite the fact that, just  like the classical equations of motion, the quantum mechanical equations of motion, either in the form of the Schr\"{o}dinger's equation or Heisenberg's equations, are time symmetric. They attribute the time-asymmetry  to the  inevitable interaction of quantum systems with macroscopic systems which lead to the collapse of the wave function, a collapse which is generally viewed as  irreversible\footnote{The causal interpretation offers a description of the measurement process which does not involve collapse. The measurement process is, by virtue of involving macroscopic devices, still viewed as irreversible but only in the statistical sense (\cite{K16} pp. 155-158).}.

Much later in 1985, the 1964 ideas concerning the values of observables in the interval between two measurements  were developed further by Albert, Aharonov and D'Amato \cite{AAA85}. They  began by noting that formula  (\ref{ABLP}) leads to the result that if the preselection operator $A$ or the postselection operator $B$ is measured  in the interval  between   the preselection and postselection measurements, the probability of obtaining the result $a$ or $b$ is $P(a)=P(b)=1$, even if $A$ and $B$  do not commute. This result, led Aharonov and his group to conclude that  in this interval  a quantum system  has  simultaneously well defined values (i.e., dispersion free values)  of both observables $A$ and $B$, irrespective of whether or not $A$ and $B$ commute  (I will comment on this below). This conclusion, the authors pointed out, appears to be contradicted by the arguments against hidden variable theories provided by Gleason \cite{G57} and Kochin and Specker \cite{KS67} (refinements of Von Neumann's impossibility proof \cite{VN1932}) in which they claim that certain sets of noncommuting observables can never be simultaneously well defined. Albert, Aharonov and D'Amato went on to show, which was the purpose of their article, that an assumption involved in the arguments of  Gleason, Kochin and Specker, an assumption  about the results of measurement of certain projection operators, was not satisfied in the interval between two measurements. This allowed them to maintain their conclusion that even two non-commuting obervables can have well defined values between two measurements.

The ideas above initiated a motivation to develop an experimental method for measuring observables  of a system between two measurements. This led  to the concept of a weak measurement,  first suggested in a 1987 article by Aharonov, Albert, Casher and Vaidmann   \cite{AACV87}. In this article, the authors claimed to  show that weak measurements can result in values of quantum observables far outside their eigenvalue spectra (we will comment on this claim below). In a later 1988 article \cite{AAV88},  Aharonov, Albert and Vaidman  developed the earlier ideas further and introduced an explicit weak measurement protocol,  a definition of a weak value of a quantum observable, and proposed a specific experiment for the measurement of the weak value of the spin of a particle, concluding that the weak value of the spin can be a 100, far outside the eigenvalue spectrum. In their mathematical analysis leading to the definition of weak value $A_w$,
\begin{equation}
A_w=\frac{\langle \phi_f|A |\phi_{in}\rangle}{\langle \phi_f|\phi_{in}\rangle},\label{WMF}
\end{equation}
 a number of assumptions were made, so that formula (\ref{WMF}) is only approximate, and it is important to keep this point in mind. These assumptions were explicitly pointed out by Duck, Stevenson and Sudarshan \cite{DSS89} in an excellent article in which they also provided a very clear and detailed mathematical analysis of the weak measurement protocol. They agreed with Aharonov et al  that weak values of quantum observables can lie far outside, even far outside,  their eigenvalue spectra.  An important practical aspect of the weak measurement protocol, pointed out by Aharonov et al at the end of their 1988 article \cite{AAV88}, is that it could be used for amplification measurements. Since this suggestion, numerous amplification experiments using the weak measurement protocol have been successfully performed. One example, is the measurement of ultrasensitive beam deflection by Dixon et al  \cite{DSJH09}.

In a later 1990 article \cite{AV90}, Aharonov and Vaidman provided a more detailed theoretical justification for the concept of a weak measurement by considering a description of quantum systems  using two wave functions. This two wave function  formalism was later called  the {\it  two-state vector formalism}, a detailed description of which can be found in reference  \cite{MME08}).  Their idea was to describe a quantum system not only by a preselected  eigenstate $| A,a\rangle$ moving forward in time, but also by  a postselected state   $| B,b\rangle$ moving backward in time. Based on this description of a quantum system, which they considered to be a time-symmetric description (though, in  my view, because the backward-in-time wave function is entirely fictitious, the time symmetry is purely mathematical, in the sense that calculated probabilities are symmetrical. In otherwords, in my view, the time symmetry does not have  physical reality), they provided a more rigorous derivation of their weak value formula (\ref{WMF}). They emphasised again, that for each individual member of the preselected and postselected ensemble a measurement of either  observable $A$ or $B$ in the intermediate interval between the measurements would yield the corresponding eigenvalue $a$ or $b$ with probability $P(a)=P(b)=1$. In fact, they asserted a stronger statement, namely, that, each member of the preselected and postselected ensemble has definite values of $A$ and $B$ in the intermediate interval, whether or not $A$ and $B$ commute, and that these definite values are $a$ and $b$, respectively.  They also once again  emphasised that  observables other than those used for preselection  and postselection can have values far outside their eigenvalue spectrum. They tried to justify the reality of weak values by arguing that there is a physical variable in the measuring device that reflects the weak value of the measured variable (\cite{AV90} p. 14) (we will comment on this below).

In reference  \cite{AAV88}, Aharonov et al suggested an experiment (hereafter the AAV-experiment) to measure the real part of the weak value of the $z$-component of a spin-half particle. Actually, for simplicity, they considered the Pauli observable  $\sigma_z$ rather the the spin observable $S_z=\frac{\hbar}{2}\sigma_z$. The idea is to couple the particles spin to its trajectory  using magnetic fields produced by pairs of Stern-Gerlach magnets. Here, the apparatus pointer  is the $z$-component of the trajectory described by its $z$-coordinate and its conjugate momentum $p'_z=p_z\pm\Delta p_z$, where  $p_z$ is the initial $z$-component of momentum, $p'_z$ is the momentum after the weak measurement and $\Delta p_z$ is the small momentum shift due to the weak measurement. The trajectory, whose bending in the $z$-direction is determined by $p'_z$, registers at a point $z$ on the detecting screen. From many such detections  the real part of the weak value of $\sigma_z$ can be determined.

A beam (ensemble) of particles moving in the positive $y$-direction (the $xy$-plane is horizontal, with the z-axis vertical forming a right-hand set) with a well defined velocity is prepared in a spin eigenstate $|\sigma_{\xi} ,+1\rangle$ (preselection), with eigenvalue +1. In this state, the spin is in the direction of  the unit vector $\hat{\xi}$  lying in the $xz$-plane at an angle $\alpha$, $90^{\circ}<\alpha<180^{\circ}$, from the positive $x$-axis. A key point of the weak measurement protocol is that the  initial value $p_z$ is very uncertain (unlike a strong von Neumann measurement which requires the initial $p_z$ to be known precisely), and is described by a broad Gaussian wave function. This means that  particles can have large values of $p_z$ in the $\pm\; z$-direction, so that by the time the beam reaches the detecting screen, it spreads considerably in the  $\pm\; z$-direction. However, because large values of $p_z$ correspond to the  Gaussian tails, only very few particles will have large values of $p_z$ and hence only a few particles will  spread widely.

 After preselection, the beam is passed through a weak magnetic field pointing in the $z$-direction. This constitutes the weak measurement. The affect is to rotate the spin-direction very slightly and to add a small $\pm\,\Delta p_z$ to the initial  $p_z$ of the particle to give a new momentum $p_z\pm\Delta p_z$, the sign depending on the initial direction of $\sigma_z$.  Note that  the weak measurement has to be sufficiently weak so that the preselected eigenstate is hardly changed.  Except for particles with initial value $p_z=0$ or with $p_z< \Delta p_z$, the initial $p_z$ dominates, so that a weak measurement revealed by a  ``spot" on the final detecting screen will  niether indicate the size nor the sign of the original value of $\sigma_z$.  For the cases $p_z=0$ or  $p_z < \Delta p_z$, the ``spot" on the detecting screen will be close to $z=0$ (relatively speaking, since the amount of  bending by the time of detection depends on the distance from the magnets the detecting screen is placed) on  the correct $\pm$ side, thus indicating a true magnitude and sign of the value of $\sigma_z$. But, even for these cases, no conclusion can be drawn from a single detection, or even from a few detections, since an exactly similar detection will be produced by a particle with opposite spin and appropriate sign and magnitude of $p_z$. For example, two particles of opposite spins with $z$-momentum values $p'_z=p_{z1}+\Delta p_z$ and $p'_z=p_{z2}-\Delta p_z$, with $p_{z2}=p_{z1}+2\Delta p_z$, after the weak measurement will produce the same spot on the detecting screen. Very many repeated measurements will reveal an average shift from which the weak value of $\sigma_z$ is determined. We have traced the motion of the particles in some, perhaps obvious, detail, since it is needed for our later comments.

The particles reaching the screen are postselected after the weak measurement by a strong magnetic field in the $+x$-direction. This strong field divides the particle beam into the $+$ and $-$ directions. Whether a particle bends into the +beam  or the -beam depends on  $\sigma_z$. The $z$-motion carrying  the weak measurement  remains unaffected. The beam (subensemble) moving in the positive $x$-direction (out of the page) is postselected and directed to the detecting screen.

If the postselected state is equal to the preselected state, the result of the weak measurement will be equal to the usual expectation value, as is obvious from Eq. (\ref{WMF}). The less orthogonal the postselected and preselected states are, the closer the measured  weak value will be to the expectation value, while the more orthogonal the states are, the farther away will be the measured weak value from the expectation value. For a sufficiently orthogonal postselected state, the weak value may lie far outside the observables eigenvlaue spectrum, as Aharonov et al emphasised. In their experiment, they showed that the measured weak value of $\sigma_z$  could be 100 (a result that we will query in our comments later). Aharonov et al also pointed out that the probability of a result decreases the more orthogonal the postselected state is, so that the probability of obtaining a result far outside the eigenvalue spectrum is very small. The physical reason for the  low probability is, that the more orthogonal the postselected state, the smaller the number of systems reaching the detecting screen, and hence, the smaller the postselected ensemble.

Though, later,  we will argue that certain choices of the postselected ensemble may give weak values never possessed  by any member of the preselected ensemble (ficitious weak values), measurements with  appropriately  chosen postselected ensembles  can lead to noval and interesting insights into the behaviour of a quantum system  between measurements. Weak measurements have also proved to be of important practical value in amplification measurements.

The first experiment using the weak measurement protocal  was performed by Ritchie et al in 1991 \cite{RSH91}. Instead of particles and Stern-Gerlach magnets,
they used an optical setup suggested by Duck et al \cite{DSS89}, in which polarisors were used to produce the preselected and postselected states and a birefringent crystal was used to perform the weak measurement. More than a decade later, a number of experiments using the weak measurement protocol to amplify weak signals were performed  (\cite{HK08} to  \cite{ ZTSN13}). In 2011 Lundeen et al performed an experiment to directly measure a wavefunction \cite{LSPSB11} following a procedure suggested by Aharonov et al \cite{AV93}. Weak measurements have also found application to quantum paradoxes. An interesting example is the experimental investigation of Hardy's paradox \cite{LS09} to  \cite{ABPRT02}.

An  experiment of particular interest using weak measurements, due to Kocsis et al \cite{KBRSMS11}, was presented in 2011. They used the weak measurement protocol, based on a theoretical proposal due to Wiseman \cite{W07},  to experimentally  determine average photon trajectories leading to interference  fringes  in a two-slit experiment. Photons emitted from a quantum dot in single photon states are divided by a 50-50 fiber beam splitter into two beams and then preselected in a diagonal linearly polarised state. The photons move forward in the $+z$-direction, while interference is along  the $\pm\,x$-directions (the $z$-axis is horizontal with the  $x$-axis vertical). The weak measurement is performed by a birefringent calcite crystal which introduces a small $k_x$-dependent phase change between the ordinary and extraordinary beams which slightly changes the linearly polarisation state to an elliptically polarised state. The photon polarisation acts as the apparatus  pointer which  indicates the $k_x$-value of the photons. The photons are postselected according to their $x$-position on the final detector (a cooled charge-coupled device). 

A quarter waveplate converts the elliptically   polarised state produced by the weak measurement to a circularly  polarised state. The conversion to a circularly polarised state allows a beam displacer to separate the ordinary and extraordinary rays in each of the two photon beams by about 2 mm vertically. This separation produces two separated interference patterns on the CCD detector  from which the $k_x$ value can be determined at each $x$-position on the CCD detector. A three lens combination with the 
middle lens movable along the $z$-axis  is positioned between the  quarter waveplate  and the beam displacer. The movable lens
images the slit system at different $z$-positions thereby changing the slits-to-CCD distance, while the CCD detector remains fixed. This allows $k_x$, hence the $k$-vector, to be measured at varies $x$-positions  along numerous $z=$ constant lines. Averaging and joining the ``dots'' produces the photon trajectories. The trajectories have just the form  predicted  by the Bohm-de Broglie  causal interpretation \cite{B52, DEBR60}.

Kocsis et al interpreted the average trajectories as the average paths of photon particles. It should be noted, however, that the Bohm-de Broglie interpretation is nonrelativistic and does not correctly describe the behaviour of the electromagnetic field. The electromagnetic field is properly described by quantum optics based on the second quantisation of Maxwell's equations. The causal interpretation of the electromagnetic field (CIEM), based on the second quantised Maxwell equations,  was developed by Kaloyerou \cite{K94}
\footnote{This development is based  on the extension of the causal interpretation to quantum fields first suggested by Bohm in paper II, p189, of reference \cite{B52} , and developed decades later in references  \cite{K85, BHK87}. Applications of CIEM to the Wheeler delayed-choice experiment \cite{WHR78} and to the Grangier-Roger-Aspect experiment \cite{GRA86} can be found in references \cite{K05,K06}, respectively.}.  In this interpretation, the electromagnetic field is viewed much like the classical concept of a field, except for additional quantum features (e.g., the field is highly nonlocal). There are no photon particles. The term ``photon'' in CIEM refers to a discrete quantum of energy $\hbar\omega$ distributed in space in the same way as any another field. At a  beam-splitter, for example, a photon is split into two beams. Therefore, as Flack and Hiley  have suggested  \cite{FH16}, the average trajectories in the Kocsis et al experiment are more correctly viewed as flow lines of the electromagnetic field.

The weak value defined in Eq. (\ref{WMF}) is a complex quantity. The physical meaning of the imaginary part is not very clear. For the case of a weak measurement of momentum, Flack and Hiley \cite{FH14} interpreted the real part of its weak value as the ordinary momentum (as defined in the causal interpretation \cite{B52}), while in the imaginary part is interpreted as the osmotic momentum (as defined in the stochastic interpretation of quantum mechanics \cite{BH89}). Also of interest, is the experimental measurement of both the real, $\mathrm{Re}(\sigma_z)_w$, and imaginary part, $\mathrm{Im}(\sigma_z)_w$, of the weak value of  a neutron's Pauli spin operator, $(\sigma_z)_w$,  by Sponar et al \cite{SDGLMTH15}. It is important to note that  $\mathrm{Im}(\sigma_z)_w$ is measured in a separate modified experiment. Sponar et al offered an interpretation of $\mathrm{Re}(\sigma_z)_w$ and  $\mathrm{Im}(\sigma_z)_w$ in terms of their affect on the total postselected state vector  representing the neutron's spin and the apparatus pointer (the apparatus pointer system consists of two paths produced in a triple Laue neutron interferometer): they asserted that $\mathrm{Re}(\sigma_z)_w$ acts as an additional phase in the state vector, while $\mathrm{Im}(\sigma_z)_w$ affects the amplitude. This interpretation does not, however, offer a physical meaning of $\mathrm{Im}(\sigma_z)_w$ in terms of a quantum observable. For example, $\mathrm{Re}(\sigma_z)_w$ represents the spin observable (after multiplication by $\hbar/2$ ), which, in the causal interpretation of the Pauli equation \cite{BST55} can be pictured in terms of a spinning particle, but there is no such immediate interpretation of $\mathrm{Im}(\sigma_z)_w$. Aharonov et al \cite{AV90} assert that the imaginary part of the weak value ``\ldots affects the distribution
of the canonical variable $q$." This again does not offer a physical meaning of the imaginary part of the weak value.

\section{Some questions and comments}

A conclusion of Aharonov et al's is that all observables, commuting or otherwise,  have definite (dispersion free) values at each instant of time, contrary to Bohr's principle of complementarity (BPC) \footnote{See reference \cite{K16} for this authors view of BPC and for further references.}. Whether or not it is believed that  all of these values can be known simultaneously,  anyone who accepts the objective reality of the wave function  will probably have held this view even prior to the work of Aharonov et al.  Certainly, those, like myself, who are supporters of the Bohm-de Broglie causal interpretation certainly believe that a quantum system simultaneously has well defined values (not necessarily eigenvalues) of all observables. Thus, in my view, the conclusion that a quantum system  has simultaneously well defined values of all observables  is entirely reasonable. Hereafter, therefore, we adopt the view that a quantum system has well defined values of all observable. What we will question, however, is what these values can be. 

Though the mathematical results are consistent with the standard quantum formalism, since they were  derived  from it, the Aharonov et al prescription for attributing definite values  to all observables  simultaneously is an extrapolation of  the usual formalism. Such an extrapolation is not unreasonable,  since any interpretation of the quantum theory, such as the Bohm-de Broglie causal interpretation, which is mathematically consistent with the standard quantum formalism, necessarily extrapolates beyond the standard formalism in terms of interpretation. A question we wish to take up below  is whether or not  Aharonov et al's extrapolation is justified.  The causal interpretation also provides a prescription for attributing definite values to all observables simultaneously, indeed, it provides  more than just a prescription, it provides rigorous formula for this purpose. We saw above, that the experimentally determined trajectories (more correctly, electromagnetic field flow lines) in the Kocsis experiment are in very good agreement with those calculated from the causal interpretation. But, we will see below that definite values of observables  attributed to systems between measurements by Aharonov et al's prescription  will sometimes differ to the the values given by the causal interpretation. The  second, perhaps more substantial, issue we want to take up concerns the reality of weak values that lie outside an observables eigenvalue spectrum.

To be specific, we ask and comment on the following questions, and note that our discussion, as in the AAV-experiment, refers to the real part of the weak value. \\  \mbox{}\\
\subsection{Question and comment 1}
In reference \cite{AAA85}, as we saw above, Aharonov et al concluded  from their formula (\ref{ABLP}) that for an ensemble preselected in state $ |A,a\rangle$ and postselected in state $|B,b\rangle$, measuring $A$ in the time interval $(t_i -t_f)$ between preselection and postselection would give the result $a$ with probability $P(a)$,  while,  if instead, $B$ is measured in the interval $(t_i -t_f)$, the result would be $b$ with probability $P(b)$. Since formula  (\ref{ABLP}) is derived from the quantum theory, this conclusion does not go beyond the usual formalism. But, in reference \cite{AV90}, Aharonov et al made the stronger assertion  that each member of an ensemble simultaneously has well defined values throughout the time interval $(t_i -t_f)$ between measurements. The latter is an extrapolation beyond the usual quantum formalism.
 {We therefore ask, {\bf``Is this extrapolation justified?"}

Every system of the preselected ensemble is in the state $| A,a\rangle$ by construction,  so that every system has the value $a$ throughout  the interval $(t_i -t_f)$. With the exception of strict Bohrians (recall the famous Wheeler assertion, ``No phenomenon is a phenomenon until it is an observed phenomenon," \cite{WHR78A}), few would argue with this view. It  is also reasonable to accept that each member of the preselected ensemble in state $|A,a \rangle$  has a definite value of $B$, though this value will vary from system to system.  There is nothing in the usual formalism that necessarily restricts  system values between measurements to eigenvalues of  $B$. Two descriptions consistent with the outcome $b$ when $B$ is measured in the interval $(t_i -t_f)$ are possible:

\begin{enumerate}
 \item[(I)]  Measurement of $B$ forces each member of the preselected ensemble into an eigenstate of $B$. Therefore, except for some chance cases where a particular system happened to have an eigenvalue of $B$, the value of $B$ for all other systems must change upon measurement of $B$. Those systems with eigenvalue $b$ are postselected.
 \item[(II)]  Aharonov et el's description: During  the interval $(t_i -t_f)$ systems in the fraction $\langle B,b|A,a\rangle$ of the preselected   ensemble in state  $| A,a\rangle$ have eigenvalue $b$ of $B$, and this fraction becomes the postselected ensemble after measurement and selection. Since, after measurement of $B$ all systems of the preselected ensemble are eigenstates of $B$ with various eigenvalues of $B$, and since an eigenstate of $B$ with an eigenvalue  other than $b$ could be postselected, then, according to Aharonov et al's description, all systems of the preselected ensemble in state $| A,a\rangle$ must have eigenvalues  of $B$ during the interval $(t_i -t_f)$.
\end{enumerate} 

The standard formalism of quantum mechanics cannot distinguish between these two descriptions. But, two arguments  suggest that description (II) is  implausible, perhaps even wrong.  

First, the causal interpretation is mathematically consistent with quantum theory and adds well defined formulae for the values of observables between measurements. Except for some special cases, such as eigenstates of an operator, the values of observables for a general state are certainly not restricted only to their eigenvalues. Thus, the predictions of the causal interpretation contradict the  assertion that well defined values of $B$ of systems in the preselected ensemble are  restricted only to eigenvalues of $B$ in the interval $(t_i-t_f)$. Morever, the value of observables incompatible with $A$ may also vary with time, and this variation is  given by the formulae of the causal interpretation, but not  by Aharonov et al's description. 

Second,  for any  system in the preselected state $ |A,a\rangle$, observable $A$ and all observables compatible with $A$ will have precise values and these values will be their eigenvalues. This is standard quantum mechanics. But if, as asserted by Aharonov et al, these systems also have well defined eigenvalues of the postselected observable $B$,  and since instead of $B$, we postselect with an observable $C$ compatible with $B$, then by Aharonov et al's reasoning, the systems in the preselected state must also have well defined values which are eigenvalues of $C$. Now, if $B$ does not commute with $A$, Aharonov et al's reasoning forces the conclusion that all systems in the preselected state  $|A,a\rangle$ will not only have well defined eigenvalues of $A$ and of all  obervables compatable  with $A$, but will also have well defined eigenvalues of $B$ and of all observables compatible with $B$. Again, this contradicts the causal interpretation predictions. Further, for nearly orthogonal preselected and postselected states, both Duck et al and Aharonov et al assert that the weak measurement can lie outside the eigenvalue spectrum. Now, if the observable being measured is compatible with either $A$ or $B$, then the weak value will contradict the assertion that systems of the preselected ensemble have values restricted to eigenvalues of this observable. 

We should emphasise that the above discussion focuses on a very specific aspect of Aharonov et al's analysis, an aspect which is not required for the correctness of the weak measurement protocol.  We also emphasise that the arguments above in no way challenge the very different assertion of Aharonov et al that an actual measurement of $A$ or $B$ in the interval $(t_i -t_f)$ results in the eigenvalues $a$ or $b$, with probabilities $P(a)=P(b)$. This conclusion holds good even for observables whose values vary with time, since the probabilities for particular eigenstate outcomes are time independent.
\subsection{Question and comment 2}
Above, it was pointed out that the more orthogonal the preselected and postselected states are, the further from the expectation value the measured weak value will be, and that weak values far outside the eigenvalue spectrum arise for nearly orthogonal preselected and postselected states. As we saw above, the latter result was graphically demonstrated by the AAV-experiment. That the apparatus pointer genuinely indicates   weak values outside the eigenvalue spectrum is not in doubt. The question is, {\bf`` Does  any member of the preselected  or postselected ensemble (or, generally,  any quantum system) actually  possess a measured weak value outside the eigenvalue spectrum?"}.

We base our answer on the AAV-experiment. The key feature in our answer is the uncertainty in the initial $p_z$ of the apparatus pointer, an essential element of a weak measurement. 

We argued above that for particles in the preselected ensemble with $p_z=0$ or  $p_z < \Delta p_z$ the final detections will indicate a correct or nearly correct value 
of $\sigma_z$. On the other hand, for particles with large $\pm\, p_z$, the large $\pm\, p_z$ masks the small $\Delta p_z$ shifts.  If the entire preselected ensemble is considered,  positive values of $p_z$, large or small, cancel (in the statistical sense, for a large enough ensemble) with corresponding negative values. This means that over many detections,  the average shift,  from which the weak value is determined,  is produced entirely by the  $\Delta p_z$ momentum shifts, shifts which correctly reflect the real value of $\sigma_z$. Thus, in this case, the weak value correctly gives the expectation value of $\sigma_z$. The same $\pm\,  p_z$ cancellations will also occur for a postselected ensemble formed from a subensemble of the preselected ensemble consisting of particles with values of $p_z$ symmetrical distributed about the central peak of the Gaussian function representing the $p_z$ distribution of the preselected ensemble. We shall call such an ensemble a ``symmteric postselected ensemble". Thus,  a symmetric postselected ensemble will also lead to a correct expectation value. Such symmetric postselected ensembles arise when the postselected state is equal to the preselected state. 

Unsymmetric postselected ensembles are formed when the postselected state is different from the preselected state. An unsymmetric ensemble is one made from a subensemble of the preselected ensemble composed of particles with a $p_z$-distribution corresponding to an off-center portion of the $p_z$-distribution of the preselected ensemble. The more the postselected state differs from the preselected state, the more unsymmetric the  postselected ensemble will be. For such an ensemble, the   $+\,p_z$ values will not balance the $-\,p_z$ values,  so that  either positive or negative $p_z$ values  will dominate, producing a contribution to the average shift in detections in addition to that due to $\Delta p_z$. Since the $\Delta p_z$ shifts reflect  the true $\sigma_z$ value, the affect of the additional $\pm\,p_z$ contribution is to distort the true value.  The weak value therefore begins to deviate from the expectation value. For  nearly orthogonal preselected and postselected states, the postselected ensemble is drawn from the tails of $p_z$-Gaussian, and hence is made up of particles with either a  very large  $+\,p_z$ or a very large $-\,p_z$, depending on which side of the central peak the particles are  drawn. In this  case, the average shift in the detections is completely dominated by the large values of $+\,p_z$ or $-\,p_z$, with the  $\Delta p_z$  making little contribution, and, hence, is also large. This large average shift in detections corresponds to a large weak value which can lie far outside the eigenvalue spectrum. We see that  this large average shift is caused by the large values of the initial $p_z$ of the apparatus pointer, which do not in anyway reflect the true value of $\sigma_z$. We conclude, that not all postselected  states lead to measured weak values that indicate  the true values possessed by systems of the preselected or postselected ensemble. In particular, weak values that lie outside the  eigenvalue spectrum are caused by large values of the initial pointer $p_z$, and not by the true values of $\sigma_z$.

A further argument against the reality of weak values outside the eigenvalue spectrum is by  comparison with the predictions of the causal interpretation. For symmetrical postselected  ensembles that give rise to weak values  of observables equal to their expectation values, the variation in the values  of these observables for each system (weak measurements cannot, of course, reveal these individual values) of either the preselected or postselected ensemble can be assumed to lie within the eigenvalue spectrum. For such cases, the values of observables calculated from the causal interpretation will be consistent with the results of a weak measurement. However, measured weak  values outside, especially far outside, the eigenvalue spectrum necessarily means that some or many individual systems of the preselected or postselected ensemble  had values also outside the eigenvalue spectrum. Such ``far out" values would  contradict values calculated from the causal interpretation.

Aharonov et al \cite{AV90} p.14 argue for the  reality of weak  values by emphasising that following  the interaction  between the system and the measuring devices ``\ldots there  is a physical variable of the measuring devices that reflects the weak value of the measured variables." That, following the interactions, there is a physical variable of the measuring device that reflects the weak value is certainly true, and, as we saw above, this variable is the momentum shift $\Delta p_z$. The problem with Aharonov et al's argument is, that, for unsymmetrical ensembles, as we argued above, this $\Delta p_z$ is masked by the higher values of $p_z$. They cannot  therefore conclude, that all weak values reflect true values possessed by individual systems. Thus, Aharonov et al's argument  justifies the reality of some weak values, but not all.

\section{Conclusion}

My comments above arose from a desire to gain an intuitive understanding of the  already  existing detailed mathematical analysis of the weak measurement protocol. A big part of the motivation  was to gain an intuitive understanding of how weak values that lie outside the eigenvalue spectrum arise. This led to the conclusion that the apparatus pointer genuinely can  register weak values outside, even far outside,  the eigenvalue spectrum, but that these values are consequence of the uncertainty in the initial $p_z$ of the apparatus pointer, and do not correspond to the true values of  observables possessed by any individual system  of either the preselected or postselected   ensemble. 

A second conclusion is, that,  an individual quantum system has  simultaneously well defined values of all observables, compatable and incompatable, in agreement  with Aharonov et al. But, contrary to  Aharonov et al,  these values are not restricted only to eigenvalues of the observables (but do lie within the eigenvalue spectra)

My comments above only suggest that  some weak values are fictatious, but by no means all. The weak measurement protocol, with appropriately chosen postselected ensembles, is, without doubt, a powerful tool allowing, for the first time, a fairly complete description of a quantum system from measurement. This allows experimental probing 
of quantum paradoxes, various foundational experiments, and even alternative interpretations of the quantum theory. Though, we have suggested that the causal interpretation can be used to discredit ``far out" weak values, with a careful choice of the postselection state, the  shoe, so to speak, is on the other foot, and weak measurements can be used investigate the causal interpretation.  We saw above that the remarkable experiment of Kocsis et al produced results in very close agreement with the predictions of the causal interpretation, thus providing strong preliminary evidence for the correctness of the  causal interpretation.  We may note that the Kocsis experiment  is an experiment in which the postselected  state produces a symmetrical postselected ensemble so that the weak values indicate true system values.

Finally, the restrictions expressed in my above comments, also do not in any way affect the use of the weak measurement protocol in``practical'' amplification experiments. Already, numerous such experiments have been performed with impressive results.

\end{document}